\documentclass{JHEP3}
\usepackage{a4wide,epsfig,amsmath,amssymb,cite}
\usepackage{scalefnt}

\newcommand{\sa}[1]{{a_{#1}}}

\newcommand{\sln}[1]{\ln^{#1} 2}
\newcommand{\sXXsX}{X_0}

\def\({\left(}
\def\){\right)}
\def\slnA{\ln2}
\def\slnB{\ln^22}
\def\slnC{\ln^32}
\def\slnD{\ln^42}
\def\slnE{\ln^52}
\def\zB{\zeta(2)}
\def\zC{\zeta(3)}
\def\zD{\zeta(4)}
\def\zE{\zeta(5)}

\def\saD{a_4}
\def\saE{a_5}
\def\lmm{\ln\frac{\mu^2}{m_t^2}}
\def\lmmB{\ln^2\frac{\mu^2}{m_t^2}}
\def\lmmC{\ln^3\frac{\mu^2}{m_t^2}}
\def\lmmD{\ln^4\frac{\mu^2}{m_t^2}}
\def\api{\frac{\alpha_s^{(n_l+1)}(\mu)}{\pi}}

\title{Four-Loop Decoupling Relations for the Strong Coupling}

\author{York Schr\"oder\\
Fakult\"at f\"ur Physik, Universit\"at Bielefeld\\
33501 Bielefeld, Germany\\
E-mail: \email{yorks@physik.uni-bielefeld.de}}

\author{Matthias Steinhauser\\
Institut f{\"u}r Theoretische Teilchenphysik,
  Universit{\"a}t Karlsruhe\\
76128 Karlsruhe, Germany\\
E-mail: \email{matthias.steinhauser@uka.de}}

\abstract{We compute the matching relation for the strong coupling constant
  within the framework of QCD up to four-loop order.
  This allows a consistent five-loop running (once the $\beta$
  function is available to this order) taking into account threshold effects.
  As a side product we obtain the effective coupling of a Higgs boson
  to gluons with five-loop accuracy.}

\preprint{BI-TP 2005/44\\SFB/CPP-05-70\\TTP05-21}


\begin{document}

%- {{{ Introduction:

\section{Introduction}

The strong coupling constant, $\alpha_s$, constitutes a fundamental 
parameter in
the Standard Model and thus its precise numerical value is very important 
for many physical predictions. An interesting property of $\alpha_s$
is its scale dependence, in particular its strong rise for low
and its small value for high energies which make perturbative
calculations within the framework of QCD possible.
The scale dependence is governed by the $\beta$ function. However, in
order to relate $\alpha_s$ at two different scales it is also necessary 
to incorporate threshold effects of heavy quarks which is achieved with
the help of the so-called matching or decoupling relations.
Thus, when specifying $\alpha_s$ it is necessary to indicate next to the 
scale also the number of active flavours.
In this paper we evaluate the decoupling relations to four-loop accuracy.
This makes it possible to 
perform a consistent running of the strong coupling evaluated at
a low scale, like, e.g., the mass of the $\tau$ lepton, to a high
scale like the $Z$ boson mass --- once the five-loop $\beta$ function
is available.

Many different techniques have been developed and applied to various
classes of Feynman diagrams. The complexity increases both with
the number of legs and the number of loops.
As far as the application of multi-loop diagrams to physical processes
is concerned the current limit are four-loop single-scale Feynman
diagrams, where either all internal particles are massless and 
one external momentum flows through the diagram 
(see, e.g., Ref.~\cite{Baikov:2004tk} for a recent publication), or
all external momenta are zero and besides massless lines there are
also particles with a common mass $M$. The latter case has been
developed in Refs.~\cite{Schroder:2002re,SturmDiss} 
and first applications can be found
in Refs.~\cite{Chetyrkin:2004fq,Schroder:2005db}. 
In this paper we consider a further
very important application: the four-loop contribution to 
the matching or decoupling relation
for the strong coupling.

The paper is organized as follows: In the next Section we 
define the decoupling constants and the theoretical framework
of our calculation.
In Section~\ref{sec::decas} we present analytical results
and discuss the numerical consequences.
In Section~\ref{sec::ggh} the connection of the decoupling constant to
the coupling of a Higgs boson to two gluons is explained and the
corresponding coupling strength is evaluated to five-loop order.
Finally, we conclude in Section~\ref{sec::concl}.
In the Appendix we present the result for the decoupling constant
parameterized in terms of the on-shell heavy quark mass.

%- }}}
%- {{{ Theoretical framework:

\section{Theoretical framework}

We consider QCD with $n_f$ active quark flavours. Furthermore it is
assumed that $n_l$ quarks are massless and $n_h$ quarks are massive, 
i.e. we have $n_f=n_l+n_h$. In practice one often has $n_h=1$, however, it
is convenient to keep a generic label for the massive quarks.

The decoupling relations relate quantities in the full and effective
theory where the latter is defined through the Lagrangian 
${\cal L}^\prime$ given by
\begin{eqnarray}
  {\cal L}^\prime\left(g_s^0,m_q^0,\xi^0;\psi_q^0,G_\mu^{0,a},c^{0,a};\zeta_i^0
  \right)
  &=&{\cal L}^{\rm QCD}\left(g_s^{0\prime},m_q^{0\prime},\xi^{0\prime};
  \psi_q^{0\prime},G_\mu^{0\prime,a},c^{0\prime,a}\right)
  \,.
  \label{eq::leff}
\end{eqnarray}
$\psi_q$, $G_\mu^a$ and $c^a$ are the fermion,
gluon and ghost fields, respectively, $m_q$ are the quark masses, $\xi$ is the
gauge parameter, and $\alpha_s = g_s^2/(4\pi)$ is the strong coupling
constant.
${\cal L}^{\rm QCD}$ is the usual QCD Lagrange density and the
effective $n_l$-flavour quantities are marked by a prime.
Eq.~(\ref{eq::leff}) states that the Lagrangian in the effective theory has
the same form as the original one with rescaled fields, masses and
coupling.
It is convenient to define the decoupling constants $\zeta_i$ in the
bare theory through
\begin{eqnarray}
  g_s^{0\prime}  =\zeta_g^0 g_s^0   \,,&\quad
  m_q^{0\prime}  =\zeta_m^0m_q^0    \,,&\quad
  \xi^{0\prime}-1=\zeta_3^0(\xi^0-1)\,,
  \nonumber\\
  \psi_q^{0\prime} =\sqrt{\zeta_2^0}\psi_q^0     \,,&\quad
  G_\mu^{0\prime,a}=\sqrt{\zeta_3^0}G_\mu^{0,a}  \,,&\quad
  c^{0\prime,a}    =\sqrt{\tilde\zeta_3^0}c^{0,a}\,.
  \label{eq::bare_dec}
\end{eqnarray}

In a next step the renormalized quantities are obtained by the usual
renormalization procedure introduced by the multiplicative
renormalization constants through~\cite{Muta}
\begin{eqnarray}
  g_s^0=\mu^{\varepsilon}Z_gg_s\,,&\qquad
  m_q^0=Z_mm_q\,,&\qquad
  \xi^0-1=Z_3(\xi-1)\,,
  \nonumber\\
  \psi_q^0=\sqrt{Z_2}\psi_q\,,&\qquad
  G_\mu^{0,a}=\sqrt{Z_3}G_\mu^a\,,&\qquad
  c^{0,a}=\sqrt{\tilde{Z}_3}c^a\,.
  \label{eq::renconst}
\end{eqnarray}

Combining Eqs.~(\ref{eq::bare_dec}) and~(\ref{eq::renconst}) leads to
renormalized decoupling constants, e.g.
\begin{eqnarray}
  \zeta_g = \frac{Z_g}{Z_g^\prime} \zeta_g^0 \,,
  \qquad
  \zeta_3 = \frac{Z_3}{Z_3^\prime} \zeta_3^0 \,,
  \qquad
  \tilde{\zeta}_3 = \frac{\tilde{Z}_3}{\tilde{Z}_3^\prime} 
  \tilde{\zeta}_3^0 \,.
  \label{eq::ren_dec}
\end{eqnarray}
Note that since we are interested in the four-loop results for
$\zeta_i$ the corresponding renormalization constants have to be known
with the same accuracy. In Ref.~\cite{Chetyrkin:2004mf} the results up to
four-loop order have
nicely been summarized (see also
Refs.~\cite{vanRitbergen:1997va,Czakon:2004bu}). 

Due to the well-known Ward identities~\cite{Muta} there are several
ways to compute the renormalization constant for the strong coupling,
$Z_g$. A convenient relation, which has the advantage that due to 
the appearance of renormalization
constants involving ghosts less diagrams contribute, is given by
\begin{eqnarray}
  Z_g &=& \frac{\tilde{Z}_1}{\tilde{Z}_3\sqrt{Z}_3}
  \,,
\end{eqnarray}
where $\tilde{Z}_1$ is the renormalization constant of the 
ghost-gluon vertex $g_s G\bar cc$. 
The same is true
for the corresponding equation for the decoupling constant,
such that one can use the relation
\begin{eqnarray}
  \zeta_g^0 &=& \frac{\tilde{\zeta}_1^0}{\tilde{\zeta}_3^0\sqrt{\zeta}_3^0}
  \,,
  \label{eq::zetag0}
\end{eqnarray}
where $\tilde{\zeta}_1^0$ denotes the decoupling constant for the 
ghost-gluon vertex.
Alternatively, one can use the renormalized objects 
$\zeta_3$, $\tilde{\zeta}_3$ from
Eq.~(\ref{eq::ren_dec}) as well as 
$\tilde{\zeta}_1 = \frac{\tilde{Z}_1}{\tilde{Z}_1^\prime}\tilde{\zeta}_1^0$ 
and then obtain $\zeta_g$ from the renormalized
version of Eq.~(\ref{eq::zetag0}).

In Refs.~\cite{Chetyrkin:1997un,Steinhauser:2002rq} formulae for the
bare decoupling constants $\zeta_i^0$ are derived which relate the
$n$-loop decoupling constants to $n$-loop vacuum integrals. In
particular, one has
\begin{eqnarray}
  \zeta_3^0 &=& 1+\Pi_G^{0h}(0)\,,
  \nonumber\\
  \tilde\zeta_3^0 &=& 1+\Pi_c^{0h}(0)\,,
  \nonumber\\
  \tilde\zeta_1^0 &=& 1+\Gamma_{G\bar cc}^{0h}(0,0)\,,
  \label{eq::bare_dec2}
\end{eqnarray}
where $\Pi_G(p^2)$ and
$\Pi_c(p^2)$ are the gluon and ghost vacuum polarizations,
respectively, and the superscript $h$ denotes the so-called hard
part which survives after setting the external momentum to zero.
Specifically, $\Pi_G(p^2)$ and $\Pi_c(p^2)$ are related to the gluon and ghost
propagators through
\begin{eqnarray}
  i\int {\rm d}x\,e^{ip\cdot x}
  \left\langle TG^{0,a\mu}(x)G^{0,b\nu}(0)\right\rangle
  \!\!&=&\!\!
  \delta^{ab}\left\{\frac{g^{\mu\nu}}{p^2\left[1+\Pi_G^0(p^2)\right]}
  +\mbox{terms proportional to $p^\mu p^\nu$}\right\}
  \,,
  \nonumber\\
  i\int {\rm d}x\,e^{ip\cdot x}
  \left\langle Tc^{0,a}(x)\bar{c}^{0,b}(0)\right\rangle
  \!\!&=&\!\!
  -\frac{\delta^{ab}}{p^2\left[1+\Pi_c^0(p^2)\right]}
  \,,
\end{eqnarray}
respectively, while
$\Gamma_{G\bar cc}^0(p,k)$ is defined through the
one-particle-irreducible (1PI) part of the amputated $G\bar cc$ Green
function as
\begin{eqnarray}
  \lefteqn{i^2\int {\rm d}x {\rm d}y\,e^{i(p\cdot x+k\cdot y)}
  \left\langle Tc^{0,a}(x)\bar c^{0,b}(0)G^{0,c\mu}(y)\right\rangle^{\rm 1PI}}
  \nonumber\\
  &\mbox{}\qquad\qquad\qquad=&
  p^\mu g_s^0\left\{-if^{abc}\left[1+\Gamma_{G\bar cc}^0(p,k)\right]
    +\mbox{other colour structures}\right\}
  \,,
\end{eqnarray}
where $p$ and $k$ are the outgoing four-momenta of $c$ and $G$, respectively,
and $f^{abc}$ are the structure constants of the QCD gauge group.
Sample four-loop diagrams for each line of Eq.~(\ref{eq::bare_dec2})
are shown in Fig.~\ref{fig::diags}(a)--(c).

\FIGURE[t]{
  \begin{tabular}{c}
  \leavevmode
  \epsfxsize=\textwidth
  \epsffile[80 380 600 500]{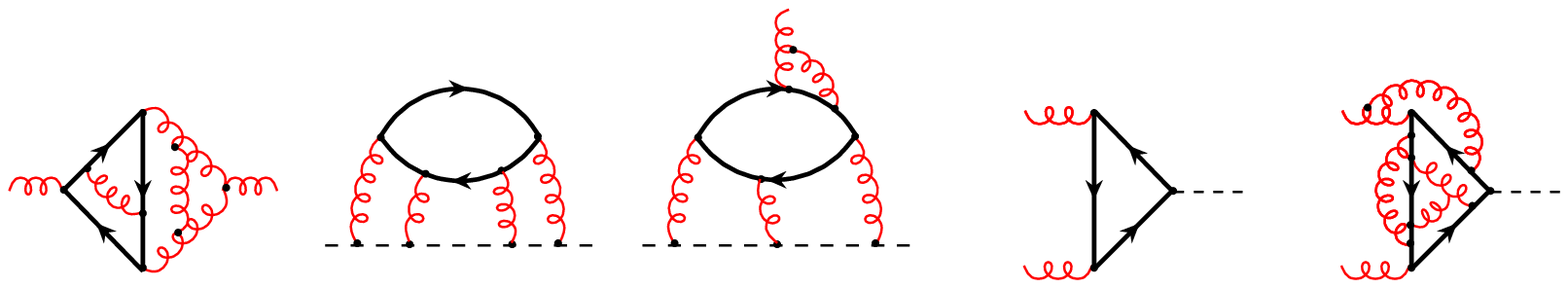}
  \\
  \mbox{}\hfill(a)\hfill(b)\hfill(c)\hfill(d)\hfill(e)\hfill\mbox{}
  \end{tabular}
  \caption{Sample diagrams for the gluon (a) and ghost (b) propagator
    and the ghost-gluon vertex (c).
    In (d) the lowest-order diagram is shown mediating the Higgs-gluon
    coupling in the Standard Model and (e) shows an example for a
    five-loop diagram contributing to the result in Eq.~(\ref{eq::c1}).
  }
  \label{fig::diags}}

From Eqs.~(\ref{eq::zetag0}),~(\ref{eq::ren_dec}) 
and~(\ref{eq::bare_dec2}) it becomes clear 
that for the calculation of $\zeta_g$ four-loop vacuum diagrams are
needed. Currently the only practical
method to express an arbitrary four-loop vacuum integral in terms of a
small set of master integrals is based on the algorithm developed in
Ref.~\cite{Laporta:2001dd}. The application to four-loop bubbles has been
discussed in Ref.~\cite{Schroder:2002re}. First physical results deal with 
moments of the photon polarization function~\cite{Chetyrkin:2004fq} and the
singlet contribution to the electroweak $\rho$
parameter~\cite{Schroder:2005db}.
The essence of the Laporta algorithm~\cite{Laporta:2001dd} is the
generation of large tables containing relations between arbitrary
integrals and the so-called master integrals. For the calculation at
hand the tables have a size of about 8~GB and contain 6~million equations.

The master integrals needed for the evaluation of $\zeta_g$ have been
computed in Ref.~\cite{Schroder:2005va}, where, however, some of the
higher order coefficients in $\epsilon$ could only be determined
numerically.

%- }}}
%- {{{ Running and decoupling for $\alpha_s$:

\section{\label{sec::decas}Running and decoupling for $\alpha_s$}

Whereas at three-loop level of the order of 1000 diagrams have to be
considered, at four loops there are almost 20000 diagrams which
contribute to the gluon and ghost propagators and the ghost-gluon
vertex. They are generated with the program {\tt
  QGRAF}~\cite{Nogueira:1991ex}. 
With the help of the packages {\tt q2e} and {\tt
  exp}~\cite{Seidensticker:1999bb,Harlander:1997zb}
the topologies and notation are adopted to the program performing the reduction
of the four-loop vacuum diagrams~\cite{Schroder:2002re}. As an output we obtain
the bare four-loop results as a linear combination
of several master integrals. All of them have been computed in
Ref.~\cite{Schroder:2005va}. 

Since at four-loop order the renormalization is quite non-trivial, let
us in the following briefly describe the procedure necessary to arrive
at a finite result.
It is convenient to build in a first step the sum of the bare
contributions to $\zeta_3^0$, $\tilde{\zeta}_3^0$ and
$\tilde{\zeta}_1^0$ and combine them immediately to
$\zeta_g^0$ according to Eq.~(\ref{eq::zetag0}). 
Already at this point the gauge parameter, $\xi$, which for the
individual pieces starts to appear at three-loop order, drops out
and hence spares us from renormalizing $\xi$.
Let us mention that due to the complexity of the intermediate
expressions, the four-loop diagrams have been evaluated for 
Feynman gauge, whereas the lower-order diagrams were computed
for general $\xi$.

In a next step it is convenient to renormalize the parameters
$\alpha_s=g_s^2/(4\pi)$ and $m_h$ applying the usual multiplicative
renormalization (cf. Eq.~(\ref{eq::renconst})). 
The corresponding counterterms have to be known up to
the three-loop order. At this point one has to apply
Eq.~(\ref{eq::ren_dec}) which requires the ratio
$Z_g/Z_g^\prime$ up to four-loop order.
In order to evaluate this ratio one has to remember that 
$Z_g^\prime$ is defined in the effective theory and thus depends on
$\alpha_s^\prime$ and $n_l$ whereas $Z_g$ depends on $\alpha_s$
and $(n_l+n_h)$. Thus it is necessary to use $\zeta_g$ up to three-loop
level in order to transform $\alpha_s^\prime$ to $\alpha_s$ where due
to the presence of the divergences in $Z_g^\prime$ also 
higher-order terms in $\epsilon$ of $\zeta_g$ have to be taken into account.

Finally one arrives at the following finite result for $(\zeta_g)^2$
which for $N_c=3$ and $n_h=1$ is given by
\begin{eqnarray}
  \zeta_g^2&=&1
  +\frac{\alpha_s^{(n_l+1)}(\mu)}{\pi}
  \left(
  -\frac{1}{6}\ln\frac{\mu^2}{m_h^2}
  \right)
  +\left(\frac{\alpha_s^{(n_l+1)}(\mu)}{\pi}\right)^2
  \left(
  \frac{11}{72} 
  -\frac{11}{24}\ln\frac{\mu^2}{m_h^2}
  +\frac{1}{36}\ln^2\frac{\mu^2}{m_h^2}
  \right)
  \nonumber\\
  &&{}+\left(\frac{\alpha_s^{(n_l+1)}(\mu)}{\pi}\right)^3
  \left[
    \frac{564731}{124416} 
    -\frac{82043}{27648}\zeta(3)
    -\frac{955}{576}\ln\frac{\mu^2}{m_h^2}
    +\frac{53}{576}\ln^2\frac{\mu^2}{m_h^2}
    -\frac{1}{216}\ln^3\frac{\mu^2}{m_h^2} 
    \right.
    \nonumber\\
    &&{}+
    \left.
    n_l\left(
    -\frac{2633}{31104}
    +\frac{67}{576}\ln\frac{\mu^2}{m_h^2} 
    -\frac{1}{36}\ln^2\frac{\mu^2}{m_h^2}
    \right)
    \right]
  +\left(\frac{\alpha_s^{(n_l+1)}(\mu)}{\pi}\right)^4
  \left[
         \frac{291716893}{6123600}
    \right.\nonumber\\&&{}\left.
       + \frac{3031309}{1306368}\slnD
       - \frac{121}{4320}\slnE
       - \frac{3031309}{217728}\zB\slnB
       + \frac{121}{432}\zB\slnC
       - \frac{2362581983}{87091200}\zC
    \right.\nonumber\\&&{}\left.
       - \frac{76940219}{2177280}\zD
       + \frac{2057}{576}\zD\slnA
       + \frac{1389}{256}\zE
       + \frac{3031309}{54432}\saD
       + \frac{121}{36}\saE
       - \frac{151369}{2177280}\sXXsX
    \right.\nonumber\\&&{}\left.
    + \left(\frac{7391699}{746496} 
    - \frac{2529743}{165888}\zeta(3)\right)
    \ln\frac{\mu^2}{m_h^2}
    + \frac{2177}{3456}\ln^2\frac{\mu^2}{m_h^2}
    - \frac{1883}{10368}\ln^3\frac{\mu^2}{m_h^2}
    + \frac{1}{1296}\ln^4\frac{\mu^2}{m_h^2}
    \right.\nonumber\\&&{}\left.
    + n_l\left(
    -\frac{4770941}{2239488} 
    + \frac{685}{124416}\sln{4}
    - \frac{685}{20736}\zeta(2)\sln{2}
    + \frac{3645913}{995328}\zeta(3) 
    \right.\right.\nonumber\\&&\left.{}
    - \frac{541549}{165888}\zeta(4)
    + \frac{115}{576}\zeta(5)
    + \frac{685}{5184}\sa{4}
    + \left(-\frac{110341}{373248} + \frac{110779}{82944}\zeta(3)
    \right)\ln\frac{\mu^2}{m_h^2}
    \right.\nonumber\\&&{}\left.
    - \frac{1483}{10368}\ln^2\frac{\mu^2}{m_h^2}
    - \frac{127}{5184}\ln^3\frac{\mu^2}{m_h^2}
    \right)
    + n_l^2\left(
    - \frac{271883}{4478976} 
    + \frac{167}{5184}\zeta(3)
    + \frac{6865}{186624}\ln\frac{\mu^2}{m_h^2}
    \right.\nonumber\\&&{}\left.\left.
    - \frac{77}{20736}\ln^2\frac{\mu^2}{m_h^2}
    + \frac{1}{324}\ln^3\frac{\mu^2}{m_h^2}
    \right)
    \right]
  + {\cal O}\left(\left(\frac{\alpha_s^{(n_l+1)}(\mu)}{\pi}\right)^5\right)
  \,,
  \label{eq::zetagana}
\end{eqnarray}
where the heavy quark mass $m_h$ is renormalized in the $\overline{\rm MS}$
scheme at the scale $\mu$. 
The corresponding expression for the on-shell mass is given in
Appendix~\ref{app::zetagos}. 
In Eq.~(\ref{eq::zetagana}), $\zeta(n)$ is Riemann's zeta function and
$\sa{n}=\mbox{Li}_n(1/2)=\sum_{k=1}^\infty 1/(2^k k^n)$. 
The constant $\sXXsX$, which is the leading coefficient of a certain 
finite four-loop master integral,
is only known numerically with the value \cite{Schroder:2005va}
\begin{eqnarray}
  \sXXsX &=& +1.808879546208334741426364595086952090\,.
\end{eqnarray}
Interestingly, in principle the number of numerical coefficients 
occurring in Eq.~(\ref{eq::zetagana}) should be three. One relation
among them can be established through the separate
renormalization of the ghost propagator while a further constant has
become available recently in analytical form~\cite{Schroder}. Thus
one remains with one coefficient which is only known numerically.

Inserting numerical values into Eq.~(\ref{eq::zetagana}) one obtains
\begin{eqnarray}
  \zeta_g^2 &\approx&1
  +0.1528\left(\frac{\alpha_s^{(n_l+1)}(m_h)}{\pi}\right)^2
  +\left(0.9721-0.0847\,n_l\right)
  \left(\frac{\alpha_s^{(n_l+1)}(m_h)}{\pi}\right)^3
  \nonumber\\&&{}
  +\left(5.1703-1.0099\,n_l-0.0220\,n_l^2\right)
  \left(\frac{\alpha_s^{(n_l+1)}(m_h)}{\pi}\right)^4
  \,.
  \label{eq::zetag}
\end{eqnarray}
It is interesting to note that the $n_l$-independent 
four-loop coefficient is relatively big as compared to the
corresponding constants at lower loop-order. However, for 
the interesting values $n_l=(3,4,5)$ one observes a big cancellation
leading to a well-defined perturbative series with coefficients
$(-0.4288,+0.7790,+1.9428)$ in front of $(\alpha_s/\pi)^4$.

We are now in a position to study the numerical impact of our result.
As an example we consider the evaluation of $\alpha_s^{(5)}(M_Z)$
from $\alpha_s^{(4)}(M_\tau)$, i.e. we apply our formalism to the
crossing of the bottom quark threshold with $n_l=4$.
In general one assumes that the value of the scale 
$\mu_b$, where the matching has to be performed, is of order $m_b$.
However, it is not determined by theory. Thus this uncertainty
contributes significantly to the error of physical predictions.
On general grounds one expects that 
while including higher order perturbative corrections
the relation
between $\alpha_s^{(4)}(M_\tau)$ and $\alpha_s^{(5)}(M_Z)$
becomes insensitive to the choice of the
matching scale. This has been demonstrated in
Refs.~\cite{Rodrigo:1997zd,Chetyrkin:1997un} 
for the three- and four-loop evolution,
respectively. In the following we want to extend the analysis to five
loops.

The procedure is as follows.
In a first step we calculate $\alpha_s^{(4)}(\mu_b)$ by exactly integrating
the equation 
\begin{eqnarray}
  \frac{\mu^2d}{d\mu^2}\,\frac{\alpha_s^{(n_f)}}{\pi}
  &=&
  \beta^{(n_f)}\left(\alpha_s^{(n_f)}\right)=
  -\sum_{i\ge0} \beta_{i}^{(n_f)}
  \left(\frac{\alpha_s^{(n_f)}}{\pi}\right)^{i+2},
  \label{eq::beta}
\end{eqnarray}
with the initial condition $\alpha_s^{(4)}(M_\tau)=0.36$.
Afterwards $\alpha_s^{(5)}(\mu_b)$ is obtained from the renormalized
version of the first equation in~(\ref{eq::bare_dec})
where we use $\zeta_g$ parameterized in terms of the on-shell mass
(cf. Eq.~(\ref{eq::zetagos})) $M_b=4.7$~GeV. 
Finally, we compute $\alpha_s^{(5)}(M_Z)$ using again
Eq.~(\ref{eq::beta}).
For consistency, $i$-loop evolution must be accompanied by $(i-1)$-loop 
matching, i.e. if we omit terms of ${\cal O}(\alpha_s^{i+2})$ on the 
right-hand side of Eq.~(\ref{eq::beta}), we need to discard those of
${\cal O}(\alpha_s^{i+1})$ in Eq.~(\ref{eq::zetagos}) at the same time.
Since the five-loop coefficient in Eq.~(\ref{eq::beta}) is not yet known we  
set $\beta_4^{(n_f)}$ to zero in our numerical analysis.

In Fig.~\ref{fig::asMZ} the result for $\alpha_s^{(5)}(M_Z)$ as a
functions $\mu_b$ is displayed for the one- to five-loop analysis.
For illustration, $\mu_b$ is varied rather extremely, by almost two orders
of magnitude.
While the leading-order result exhibits a strong logarithmic behaviour, the
analysis is gradually getting more stable as we go to higher orders.
The five-loop curve is almost flat for $\mu_b \ge 1$~GeV and
demonstrates an even more stable behaviour than the four-loop analysis
of Ref.~\cite{Chetyrkin:1997un}. It should be noted that around $\mu_b \approx
1$~GeV both the three-, four- and five-loop curves show a strong
variation which can be interpreted as a sign for the breakdown of
perturbation theory.
Besides the $\mu_b$ dependence of $\alpha_s^{(5)}(M_Z)$, also its absolute 
normalization is significantly affected by the higher orders.
At the central matching scale $\mu_b=M_b$, we encounter a rapid
convergence behaviour.

\FIGURE[t]{
  \epsfxsize=\textwidth
  \epsfig{figure=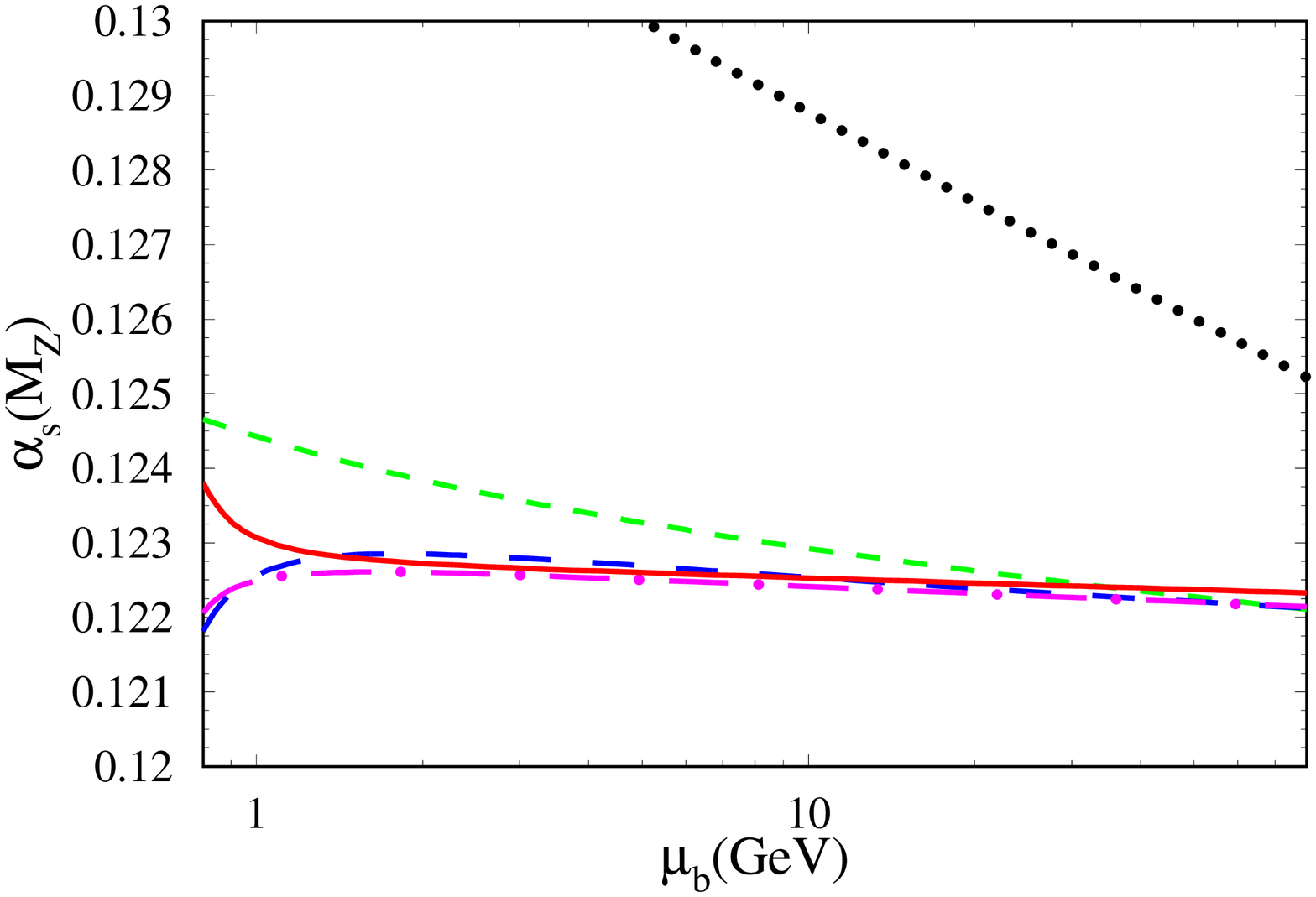,width=36em}
  \caption{$\mu_b$ dependence of $\alpha_s^{(5)}(M_Z)$ calculated from
    $\alpha_s^{(4)}(M_\tau)=0.36$ and $M_b=4.7$~GeV. The procedure is
    described in the text. The dotted, short-dashed, long-dashed and
    dash-dotted line corresponds to 
    one- to four-loop running. The solid curve 
    includes the effect of the new four-loop matching term.
  }
  \label{fig::asMZ}}

%- }}}
%- {{{ Effective coupling between a Higgs boson and gluons:

\section{\label{sec::ggh}
Effective coupling between a Higgs boson and gluons}

In this Section we want to discuss the relation between $\zeta_g$ and
the coupling of a scalar Higgs boson to gluons. Due to the fact that
gluons are massless, there is no coupling at tree-level. At one-loop
order the $HGG$ coupling is mediated via a top-quark loop depicted in
Fig.~\ref{fig::diags}(d).

For an intermediate-mass Higgs boson which formally obeys the relation
$M_H\ll m_t$ it is possible to construct an effective Lagrangian of
the form
\begin{eqnarray}
  {\cal L}_{\rm eff} &=& -\frac{H^0}{v^0} C_1 {\cal O}_1
  \,,
  \label{eq::Leff}
\end{eqnarray}
with the effective operator
\begin{eqnarray}
  {\cal O}_1 &=& \left(G^{a}_{\mu\nu}\right)^2\,,
\end{eqnarray}
where $G_{\mu\nu}^a$ is the colour field strength.
The coefficient function $C_1$ incorporates the contribution from the
top-quark loops. At one-loop order it is easy to see that 
the contribution from the triangle diagrams
can be obtained through the derivative of the one-loop diagram
for $\Pi_G^0$ with respect to the top-quark mass.
However, at higher-loop orders this simple picture does not hold 
anymore and the relation between the $HGG$ diagrams and derivatives of the 
two-point functions containing a top-quark loop gets more involved.
In Ref.~\cite{Chetyrkin:1997un} an all-order low-energy theorem has
been derived which establishes such a relation and which has a surprisingly
simple form (for definiteness we specify to the top-quark in this Section):
\begin{eqnarray}
  C_1 &=& -\frac{1}{2}\,\frac{m_t^2\partial}{\partial m_t^2}\,
  \ln\zeta_g^2
  \,.
  \label{eq::c1let1}
\end{eqnarray}
An appealing feature of Eq.~(\ref{eq::c1let1}) is that at a given
order in $\alpha_s$
only the logarithmic contributions of $\zeta_g$ are needed for the 
calculation of $C_1$ at the same order. Thus, from our calculation
we can reconstruct the five-loop logarithms of $\zeta_g$ from lower-order 
terms and the $\beta$ and $\gamma_m$ functions governing the running 
of $\alpha_s$ and the top-quark mass, respectively.
This leads to the following result, at $N_c=3$ and $n_h=1$, 

{
\begin{eqnarray}
  C_1 &=&
  -\frac{1}{12}\,\api
  \Bigg\{
  1 
  + \api
  \(
  \frac{11}{4} 
  - \frac{1}{6} \ln\frac{\mu^2}{m_t^2}
  \)
  \nonumber\\&&{}
+ \(\api\)^2
  \Bigg[
    \frac{2821}{288} 
    - \frac{3}{16} \ln\frac{\mu^2}{m_t^2}
    + \frac{1}{36} \ln^2\frac{\mu^2}{m_t^2}
    + n_l\left(
    -\frac{67}{96} 
    + \frac{1}{3} \ln\frac{\mu^2}{m_t^2}
    \right)
    \Bigg]
  \nonumber\\&&{}
+ \(\api\)^3
  \Bigg[
    -\frac{4004351}{62208} 
    + \frac{1305893}{13824}\zeta(3)
    - \frac{859}{288} \ln\frac{\mu^2}{m_t^2}
    + \frac{431}{144} \ln^2\frac{\mu^2}{m_t^2}
    - \frac{1}{216} \ln^3\frac{\mu^2}{m_t^2}
    \nonumber\\&&{}
    +  n_l \left(
    \frac{115607}{62208} 
    - \frac{110779}{13824}\zeta(3)
    + \frac{641}{432} \ln\frac{\mu^2}{m_t^2}
    + \frac{151}{288} \ln^2\frac{\mu^2}{m_t^2}
    \right) 
    \nonumber\\&&{}
    + n_l^2 \left(
    - \frac{6865}{31104} 
    + \frac{77}{1728} \ln\frac{\mu^2}{m_t^2} 
    - \frac{1}{18} \ln^2\frac{\mu^2}{m_t^2}
    \right)
    \Bigg]
  \nonumber\\&&{}
+ \(\api\)^4
  \Bigg[
       - \frac{69820734619}{27993600}
       - \frac{39407017}{373248}\slnD
       + \frac{11011}{8640}\slnE
       + \frac{39407017}{62208}\zB\slnB
\nonumber\\&&{}
       - \frac{11011}{864}\zB\slnC
       + \frac{27642438179}{24883200}\zC
       + \frac{996205247}{622080}\zD
       - \frac{187187}{1152}\zD\slnA
       - \frac{894391}{4608}\zE
\nonumber\\&&{}
       - \frac{39407017}{15552}\saD
       - \frac{11011}{72}\saE
       + \frac{1967797}{622080}\sXXsX
\nonumber\\&&{}
- \( \frac{1276661933}{1492992} - \frac{226222121}{331776}\zC \)\lmm 
+ \frac{33517}{1728}\lmmB 
+ \frac{140357}{20736}\lmmC 
+ \frac{1}{1296}\lmmD 
\nonumber\\&&{}
+  n_l \(
         \frac{58259821853}{195955200}
       + \frac{3896297}{580608}\slnD
       - \frac{121}{1440}\slnE
       - \frac{3896297}{96768}\zB\slnB
       + \frac{121}{144}\zB\slnC
\right.\nonumber\\&&{}\left.
       - \frac{74306021071}{348364800}\zC
       + \frac{141211087}{3870720}\zD
       + \frac{2057}{192}\zD\slnA
       - \frac{20227}{2304}\zE
       + \frac{3896297}{24192}\saD
       + \frac{121}{12}\saE
\right.\nonumber\\&&{}\left.
       - \frac{151369}{725760}\sXXsX
+ \( \frac{23250409}{186624} - \frac{8736121}{82944}\zC \)\lmm 
+ \frac{569}{2304}\lmmB 
+ \frac{2551}{2592}\lmmC 
\)
\nonumber\\&&{}
+  n_l^2 \(
-\frac{33014371}{8957952} 
+ \frac{685}{41472}\slnD 
- \frac{685}{6912}\zB\slnB 
+ \frac{970259}{110592}\zC 
- \frac{518509}{55296}\zD 
\right.\nonumber\\&&{}\left.
+ \frac{115}{192}\zE
+ \frac{685}{1728}\saD 
- \( \frac{1107181}{186624} - \frac{28297}{9216}\zC \)\lmm 
- \frac{1729}{13824}\lmmB 
- \frac{1205}{5184}\lmmC 
\) 
\nonumber\\&&{}
+  n_l^3 \(
-\frac{255947}{1492992} 
+ \frac{5}{64}\zC
+ \frac{481}{5184}\lmm 
- \frac{77}{6912}\lmmB 
+ \frac{1}{108}\lmmC
\) 
\nonumber\\&&{}
+ 6\left(\beta_4^{(n_l)} - \beta_4^{(n_l+1)}\right)
    \Bigg]
  + {\cal O}\(\(\api\)^5\)
  \Bigg\}
  \,,
  \label{eq::c1}
\end{eqnarray}
}

\noindent
with $m_t$ being the $\overline{\rm MS}$ top-quark mass 
renormalized at the scale $\mu$.
Note the appearance of the flavour-dependent part of
$\beta_4$ in the five-loop contribution,
whereas the corresponding coefficient from the anomalous mass
dimension does not appear.
We want to stress that the term of order $\alpha_s^5$ covers the
contributions from five-loop diagrams like the one in
Fig.~\ref{fig::diags}(e). 

Evaluating Eq.~(\ref{eq::c1}) numerically leads to
\begin{eqnarray}
  C_1 &\approx&
  -\frac{1}{12}\,\frac{\alpha_s^{(n_l+1)}(m_t)}{\pi}
  \Bigg[1
    + 2.7500\,
    \frac{\alpha_s^{(n_l+1)}(m_t)}{\pi}
    + \left(9.7951 - 0.6979\,n_l\right) 
    \left(\frac{\alpha_s^{(n_l+1)}(m_t)}{\pi}\right)^2
    \nonumber\\&&\mbox{}
    + \left(49.1827 - 7.7743\,n_l - 0.2207\,n_l^2 \right)
    \left(\frac{\alpha_s^{(n_l+1)}(m_t)}{\pi}\right)^3
    \nonumber\\&&\mbox{}
    + \left(-662.5065 + 137.6005\,n_l -2.5367\,n_l^2 -0.0775\,n_l^3
    + 6\left(\beta_4^{(n_l)} - \beta_4^{(n_l+1)}\right)
    \right)
    \nonumber\\&&\mbox{}
    \times\left(\frac{\alpha_s^{(n_l+1)}(m_t)}{\pi}\right)^4
    \Bigg]
  \,.
\end{eqnarray}
Again one observes large cancellations between the $n_l^0$ and $n_l^1$
term in the five-loop contribution to $C_1$. 

Note that the result of Eq.~(\ref{eq::c1}) constitutes a building
block for the N$^4$LO calculation to the Higgs boson production and
decay in the two-gluon channel, for which the complete answer
currently is certainly out of range. Still, the five-loop result for
$C_1$ constitutes a high-order result in perturbative QCD which is of
theoretical interest by itself.

%- }}}
%- {{{ Conclusions:

\section{\label{sec::concl}Conclusions}

In this paper the decoupling constant of the strong coupling
is presented to four-loop order. This constitutes a fundamental
quantity of QCD and is one of the very few known to such a high order.
The decoupling constant is necessary for performing a consistent running
of $\alpha_s$ with five-loop accuracy including important effects from
the crossing of quark thresholds. The calculation has been performed
analytically, and the main result can be found in
Eq.~(\ref{eq::zetagana}).
With the help of a low-energy theorem 
it is possible to derive the five-loop result
for the effective coupling of the Higgs boson to gluons, which
constitutes a building block in the corresponding production and decay
processes.

We want to mention that the result for $\zeta_g^2$ in
Eq.~(\ref{eq::zetagana}) has been obtained independently in 
Ref.~\cite{ChetDec}. Except for {\tt QGRAF}, which is used for the
generation of the diagrams, there is no common code. Even the master
integrals have meanwhile been computed independently~\cite{ChetMas}
and for the renormalization a different procedure has been chosen.

%- }}}
%- {{{ Acknowledgements:

\vspace*{1em}

\noindent
{\bf Acknowledgements}\\
This work was supported by the SFB/TR 9.
We would like to thank the authors of Ref.~\cite{ChetDec} for
communicating their result before publication.

%- }}}
%- {{{ Appendix:

\begin{appendix}

\section{\label{app::zetagos}Results for $\zeta_g^{\rm OS}$}

Replacing in Eq.~(\ref{eq::zetagana}) the $\overline{\rm MS}$ mass
$m_h$ by the pole mass $M_h$ using the three-loop
approximation~\cite{Chetyrkin:1999ys,Chetyrkin:1999qi,Melnikov:2000qh} one gets
\def\lmmOS{\ln\frac{\mu^2}{M_h^2}}
\def\lmmOSB{\ln^2\frac{\mu^2}{M_h^2}}
\def\lmmOSC{\ln^3\frac{\mu^2}{M_h^2}}
\def\lmmOSD{\ln^4\frac{\mu^2}{M_h^2}}
\def\apiOS{\frac{\alpha_s^{(n_l+1)}(M_h)}{\pi}}

\begin{eqnarray}
\(\zeta_g^{\rm OS}\)^2 &=& 
1
+ \api \(
- \frac{1}{6}\lmmOS
\)
+ \(\api\)^2 \(
- \frac{7}{24} 
- \frac{19}{24}\lmmOS
+ \frac{1}{36}\lmmOSB
\)
\nonumber\\&&{}
+\(\api\)^3
\left[
-\frac{58933}{124416}
-\frac{2}{3}\zeta(2)
-\frac{2}{9}\zeta(2)\ln2
-\frac{80507}{27648}\zeta(3)
-\frac{8521}{1728}\lmmOS
\right.\nonumber\\&&{}\left.
-\frac{131}{576}\lmmOSB
-\frac{1}{216}\lmmOSC 
+n_l\(
\frac{2479}{31104}
+\frac{1}{9}\zeta(2)
+\frac{409}{1728}\lmmOS 
\)
\right]
\nonumber\\&&{}
+\(\api\)^4
\Bigg[
       - \frac{141841753}{24494400}
       + \frac{3179149}{1306368}\slnD
       - \frac{121}{4320}\slnE
       - \frac{697121}{19440}\zB
\nonumber\\&&{}
       + \frac{1027}{162}\zB\slnA
       - \frac{2913037}{217728}\zB\slnB
       + \frac{121}{432}\zB\slnC
       - \frac{2408412383}{87091200}\zC
\nonumber\\&&{}
       + \frac{1439}{216}\zC\zB
       - \frac{71102219}{2177280}\zD
       + \frac{2057}{576}\zD\slnA
       + \frac{49309}{20736}\zE
\nonumber\\&&{}
       + \frac{3179149}{54432}\saD
       + \frac{121}{36}\saE
       - \frac{151369}{2177280}\sXXsX
\nonumber\\&&{}
- \( \frac{19696909}{746496} + \frac{29}{9}\zB 
  + \frac{29}{27}\zB\slnA + \frac{2439119}{165888}\zC \)\lmmOS
- \frac{7693}{1152}\lmmOSB 
\nonumber\\&&{}
- \frac{8371}{10368}\lmmOSC 
+ \frac{1}{1296}\lmmOSD
+ n_l\(
\frac{1773073}{746496} 
+ \frac{173}{124416}\slnD 
+ \frac{557}{162}\zB 
\right.\nonumber\\&&\left.{}
+ \frac{22}{81}\zB\slnA 
- \frac{1709}{20736}\zB\slnB
+ \frac{4756441}{995328}\zC 
- \frac{697709}{165888}\zD 
+ \frac{115}{576}\zE
+ \frac{173}{5184}\saD 
\right.\nonumber\\&&\left.{}
+ \( \frac{1110443}{373248} + \frac{41}{54}\zB 
  + \frac{2}{27}\zB\slnA + \frac{132283}{82944}\zC \)\lmmOS 
+ \frac{6661}{10368}\lmmOSB 
\right.\nonumber\\&&\left.{}
+ \frac{107}{1728}\lmmOSC 
\) 
+ n_l^2\(
-\frac{140825}{1492992} 
- \frac{13}{162}\zB 
- \frac{19}{1728}\zC
\right.\nonumber\\&&\left.{}
- \( \frac{1679}{186624} + \frac{1}{27}\zB \)\lmmOS 
- \frac{493}{20736}\lmmOSB 
\) 
\Bigg]
\nonumber\\
&\approx&
1
-0.2917 \(\apiOS\)^2
+\(-5.3239+0.2625\,n_l\) \(\apiOS\)^3
\nonumber\\&&{}
+\(-85.8750+9.6923\,n_l-0.2395\,n_l^2\) \(\apiOS\)^4
\,.
\label{eq::zetagos}
\end{eqnarray}

\end{appendix}

%- }}}
%- {{{ bibliography:

%- }}}


\begin{thebibliography}{99}

\bibitem{Baikov:2004tk}
  P.~A.~Baikov, K.~G.~Chetyrkin and J.~H.~K\"uhn,
  %``Strange quark mass from tau lepton decays with O(alpha(s**3)) accuracy,''
  Phys.\ Rev.\ Lett.\  {\bf 95} (2005) 012003
  [hep-ph/0412350].
  %%CITATION = HEP-PH 0412350;%%

%\cite{Schroder:2002re}
\bibitem{Schroder:2002re}
  Y.~Schr\"oder,
  %``Automatic reduction of four-loop bubbles,''
  Nucl.\ Phys.\ Proc.\ Suppl.\  {\bf 116} (2003) 402
  [hep-ph/0211288].
  %%CITATION = HEP-PH 0211288;%%

\bibitem{SturmDiss}
  C. Sturm, PhD thesis (June 2005), Karlsruhe Univ., Germany (unpublished).

%\cite{Chetyrkin:2004fq}
\bibitem{Chetyrkin:2004fq}
  K.~G.~Chetyrkin, J.~H.~K\"uhn, P.~Mastrolia and C.~Sturm,
  %``Heavy-quark vacuum polarization: First two moments of the O(alpha(s)**3
  %n(f)**2) contribution,''
  Eur.\ Phys.\ J.\ C {\bf 40} (2005) 361
  [hep-ph/0412055].
  %%CITATION = HEP-PH 0412055;%%

%\cite{Schroder:2005db}
\bibitem{Schroder:2005db}
  Y.~Schr\"oder and M.~Steinhauser,
  %``Four-loop singlet contribution to the rho parameter,''
  Phys.\ Lett.\ B {\bf 622} (2005) 124
  [hep-ph/0504055].
  %%CITATION = HEP-PH 0504055;%%

\bibitem{Muta}
  T. Muta, {\it Foundations of Quantum Chromodynamics}, World
  Scientific, Singapore, 1987.

%\cite{Chetyrkin:2004mf}
\bibitem{Chetyrkin:2004mf}
  K.~G.~Chetyrkin,
  %``Four-loop renormalization of QCD: Full set of renormalization constants
  %and anomalous dimensions,''
  Nucl.\ Phys.\ B {\bf 710} (2005) 499
  [hep-ph/0405193].
  %%CITATION = HEP-PH 0405193;%%

%\cite{vanRitbergen:1997va}
\bibitem{vanRitbergen:1997va}
  T.~van Ritbergen, J.~A.~M.~Vermaseren and S.~A.~Larin,
  %``The four-loop beta function in quantum chromodynamics,''
  Phys.\ Lett.\ B {\bf 400} (1997) 379
  [hep-ph/9701390].
  %%CITATION = HEP-PH 9701390;%%

%\cite{Czakon:2004bu}
\bibitem{Czakon:2004bu}
  M.~Czakon,
  %``The four-loop QCD beta-function and anomalous dimensions,''
  Nucl.\ Phys.\ B {\bf 710} (2005) 485
  [hep-ph/0411261].
  %%CITATION = HEP-PH 0411261;%%

%\cite{Schroder:2005va}
\bibitem{Chetyrkin:1997un}
  K.~G.~Chetyrkin, B.~A.~Kniehl and M.~Steinhauser,
  %``Decoupling relations to O(alpha(s)**3) and their connection to  low-energy
  %theorems,''
  Nucl.\ Phys.\ B {\bf 510} (1998) 61
  [hep-ph/9708255].
  %%CITATION = HEP-PH 9708255;%%

%\cite{Steinhauser:2002rq}
\bibitem{Steinhauser:2002rq}
  M.~Steinhauser,
  %``Results and techniques of multi-loop calculations,''
  Phys.\ Rept.\  {\bf 364} (2002) 247
  [hep-ph/0201075].
  %%CITATION = HEP-PH 0201075;%%

%\cite{Laporta:2001dd}
\bibitem{Laporta:2001dd}
  S.~Laporta,
  %``High-precision calculation of multi-loop Feynman integrals by  difference
  %equations,''
  Int.\ J.\ Mod.\ Phys.\ A {\bf 15} (2000) 5087
  [hep-ph/0102033].
  %%CITATION = HEP-PH 0102033;%%

%\cite{Rodrigo:1997zd}
\bibitem{Schroder:2005va}
  Y.~Schr\"oder and A.~Vuorinen,
  %``High-precision epsilon expansions of single-mass-scale four-loop vacuum
  %bubbles,''
  JHEP {\bf 0506} (2005) 051
  [hep-ph/0503209].
  %%CITATION = HEP-PH 0503209;%%

%\cite{Nogueira:1991ex}
\bibitem{Nogueira:1991ex}
  P.~Nogueira,
  %``Automatic Feynman graph generation,''
  J.\ Comput.\ Phys.\  {\bf 105} (1993) 279.
  %%CITATION = JCTPA,105,279;%%

%\cite{Seidensticker:1999bb}
\bibitem{Seidensticker:1999bb}
  T.~Seidensticker,
  %``Automatic application of successive asymptotic expansions of
  %Feynman
  %diagrams,''
  hep-ph/9905298.
  %%CITATION = HEP-PH 9905298;%%

%\cite{Harlander:1997zb}
\bibitem{Harlander:1997zb}
  R.~Harlander, T.~Seidensticker and M.~Steinhauser,
  %``Complete corrections of O(alpha alpha(s)) to the decay of the Z
  %boson  into
  %bottom quarks,''
  Phys.\ Lett.\ B {\bf 426} (1998) 125
  [hep-ph/9712228].
  %%CITATION = HEP-PH 9712228;%%

\bibitem{Schroder}
  Y.~Schr\"oder, in preparation.

%\cite{Bernreuther:1981sg}
\bibitem{Rodrigo:1997zd}
  G.~Rodrigo, A.~Pich and A.~Santamaria,
  %``alpha(s)(m(Z)) from tau decays with matching conditions at three loops,''
  Phys.\ Lett.\ B {\bf 424} (1998) 367
  [hep-ph/9707474].
  %%CITATION = HEP-PH 9707474;%%

\bibitem{ChetDec}
  K.~G.~Chetyrkin, J.~H.~K\"uhn and C.~Sturm, hep-ph/0512060.
%  Report No.: TTP05-XX, hep-ph/0510XXX

\bibitem{ChetMas}
  K.~G.~Chetyrkin, M.~Faisst, C.~Sturm and M. Tentukov, in preparation.
%  Report No.: TTP05-XX, hep-ph/0510XXX

%\cite{Chetyrkin:1999ys}
\bibitem{Chetyrkin:1999ys}
  K.~G.~Chetyrkin and M.~Steinhauser,
  %``Short distance mass of a heavy quark at order alpha(s)**3,''
  Phys.\ Rev.\ Lett.\  {\bf 83} (1999) 4001
  [hep-ph/9907509].
  %%CITATION = HEP-PH 9907509;%%

%\cite{Chetyrkin:1999qi}
\bibitem{Chetyrkin:1999qi}
  K.~G.~Chetyrkin and M.~Steinhauser,
  %``The relation between the MS-bar and the on-shell quark mass at order
  %alpha(s)**3,''
  Nucl.\ Phys.\ B {\bf 573} (2000) 617
  [hep-ph/9911434].
  %%CITATION = HEP-PH 9911434;%%

%\cite{Melnikov:2000qh}
\bibitem{Melnikov:2000qh}
  K.~Melnikov and T.~v.~Ritbergen,
  %``The three-loop relation between the MS-bar and the pole quark masses,''
  Phys.\ Lett.\ B {\bf 482} (2000) 99
  [hep-ph/9912391].
  %%CITATION = HEP-PH 9912391;%%

\end{thebibliography}
\end{document}